# Evolutionary games: natural selection of strategies

G. P. Karev

National Center for Biotechnology Information, National Institutes of Health - Bldg. 38A, 8600 Rockville Pike, Bethesda, MD 20894, USA. Email: karev@ncbi.nlm.nih.gov.

**Abstract**

In this paper, I model and study the process of natural selection between all possible mixed strategies in classical two-player two-strategy games. I derive and solve an equation that is a natural generalization of the Taylor-Jonker replicator equation that describes dynamics of pure strategy frequencies. I then investigate the evolution of not only frequencies of pure strategies but also of total distribution of mixed strategies. I show that the process of natural selection of strategies for all games obeys the dynamical Principle of minimum of information gain. It is also shown a principle difference between mixed-strategies Hawk-Dove (HD) game and all other games (Prisoner's Dilemma, Harmony and Stag-Hunt games). Mathematically, the limit distribution of strategies is non-singular and the information gain tends to a finite value for HD-game, in contrast with all other games. Biologically, the process of natural selection in HD-game follows non-Darwinian "selection of everybody" but for all other games we observe Darwinian "selection of the fittest".

## Introduction. Problem formulation

Mathematical game theory was initially developed for economic and social problems to make predictions about how different behavioral strategies can affect individuals' behavior through affecting the predicted "payoff" that corresponds to each strategy. Evolutionary game theory (EGT) originated as its application to biological problems, as, per the famous dictum of T.Dobzhansky "Nothing in biology makes sense except in the light of evolution" (Dobzhansky, 1973; see also an interesting discussion of this dictum in Griffiths, 2009). It looks at strategies that species, or individuals within species can employ to increase their payoff, which in the case of EGT is their fitness.

First explicit applications of mathematical game theory to evolutionary biology were given in (Levontin, 1961) and (Maynard Smith and G. Price, 1973), as was formalized in a seminal monograph by Maynard Smith (1982). EGT further found interesting applications in social sciences, where evolution is not biological but economical or cultural evolution; EGT provides appropriate assumptions and mathematical tools for investigation of corresponding problems. Within EGT, strategies are seen as inherited programs that control the individuals' behavior. "…The word strategy could be replaced by the word phenotype; for example, a strategy could be the growth form of a plant, or the age at first reproduction, or the relative numbers of sons and daughters produced by a parent" (Maynard Smith 1982, ch.1). Understanding mechanisms underlying strategy selection can allow making predictions about evolutionary trajectories, deepening our understanding of the underlying biology.

Let us consider a specific example. Assume a large population of individuals (of the same species) that can adopt two different strategies, which we denote as 0 and 1. Strategy $i$ player receives payoff $G_{ij}$ when playing against strategy $j$ player; $G = (G_{ij})$, $i, j = 0,1$ is a matrix of expected payoffs. Within the frameworks of EGT, the payoff is considered as individual's fitness. Below we will call the 0-strategy players "cooperators" and 1-strategy players "defectors", keeping in line with the classical games of "prisoner's dilemma" and "hawk-dove" (see Hofbauer and Sigmund 1998, Broom and Rychtár, 2013 for details). These games will be discussed in greater detail in the following sections.

The central concept suggested in (Maynard Smith and Price, 1973) for game analysis was the concept of an evolutionarily stable, or uninvadable strategy (ESS). An ESS is a strategy such that, if all the members of a population adopt it, then no mutant strategy could invade the population under the influence of natural selection.

Later a dynamical approach for game analysis known as replicator dynamics was offered by Taylor and Jonker (1978); for details and further development of replicator dynamics see, e.g., (Hofbauer and Sigmund 1998; Webb 2007). In the simplest problem formulation, it is assumed that every individual can adopt only a single pure strategy; the population is divided into two subpopulations with frequencies $x(t)$ and $1-x(t)$ such that all individuals in the first subpopulation adopt strategy 1 (defectors) and in the second subpopulation they adopt strategy 0 (cooperators). Then the dynamics of the frequency $x(t)$ is described by the Taylor-Jonker replicator equation

$$\frac{dx}{dt} = x(1-x)((A+B)x - A), \tag{0.1}$$

where

$$A = G_{00} - G_{10}, \quad B = G_{11} - G_{01}. \tag{0.2}$$

This way the dynamics of strategy frequencies depend on only two quantities, $A$ and $B$, instead of the original four elements of the payoff matrix.

Equation (0.1) always has equilibria $x=0$ and $x=1$ and may have an additional equilibrium $x^* = \frac{A}{A+B}$ if $0 < \frac{A}{A+B} < 1$.

Replicator equation (0.1) describes four well-known classes of games depending on the signs of $A$ and $B$. If $A<0$ and $B>0$, then $x=1$ is a stable equilibrium; the corresponding game is Prisoner's Dilemma (PD), where defectors dominate over cooperators. If $A>0$ and $B<0$, then $x=0$ is a stable equilibrium; the corresponding game is Harmony (H), where cooperators dominate over defectors. In both cases, there are no other stables states. These two games are identical up to reordering of the strategies.

If both $A$ and $B$ are positive, then the game becomes Stag Hunt (SH) or coordination game. It is characterized by bi-stability, when both equilibria $x=0$ and $x=1$ are stable and their areas of attraction are divided by the unstable equilibrium $x^* = \frac{A}{A+B}$.

Finally, if both $A$ and $B$ are negative, then there exists internal (polymorphic) stable equilibrium $x^* = A/(A+B)$ and equilibria $x=0$ and $x=1$ are unstable; this is the Hawk–Dove (HD) game, also known as snowdrift or the game of chicken.

The main results on stability properties of replicator dynamics can be found in (Cressman, 1992; Hoffbauer and Zigmund, 1998).

In addition to the two described approaches to studying evolutionary games, namely, studying evolutionarily stable strategies and modeling the dynamics of strategy frequencies, we develop here a third approach, which generalizes the second one. Specifically, we model and study the process of natural selection between all possible mixed strategies in the population. In what follows we assume that every individual can adhere not only to one "pure" strategy but can adopt each strategy with its own (hereditary) probabilities, i.e. keep a mixed strategy. We derive and solve an equation that is a natural generalization of the Taylor-Jonker replicator equation (0.1). We then investigate the evolution of not only frequencies of pure strategies but of total distribution of mixed strategies.

## The model and the main equations

Let $l(t,\alpha)$ be the set of all individuals that adopt strategy 1 (defection) with probability $\alpha$ and strategy 0 (cooperation) with probability $1-\alpha$; we refer to the subpopulation $l(t,\alpha)$ as $\alpha$-clone. Then $l(t,0)$ consists only of "cooperators" and $l(t,1)$ consists only of "defectors"; in the simplest case, the total population consists of these two clones only. In our case, there exists an indefinite number of mixed strategies, which can be adopted by individuals from the population, and each mixed strategy is characterized by its own value of the parameter $\alpha, 0 \leq \alpha \leq 1$. In what follows we will denote the density of $\alpha$-clone also by $l(t,\alpha)$.

Our aim is to trace the natural selection of mixed strategies in process of "honest competition"; this process is described by the evolution of distribution of the parameter $\alpha$.

In order to formulate a mathematical model, let us introduce the necessary notations and exact definitions. Let $P(0,\alpha)$ be the pdf (probability density function) of the initial distribution of the parameter $\alpha$. Let $S_0$ be the support of $P(0,\alpha)$, i.e. $S_0 = \{\alpha : P(0,\alpha) > 0\}$. For example, if $S_0$ consist of only two points, $S_0 = \{0,1\}$, then the population is composed of only two clones, one of which consists of cooperators and another of defectors. If all mixed strategies are possible in the population, then $S_0$ coincides with the unit interval, $S_0 = [0,1]$; it is the case of our primary interest. In what follows we always assume that the points $\alpha = 0$ and $\alpha = 1$ belong to $S_0$; it means that individuals that adopt pure strategies are present in the population at initial time moment.

The total population size at $t$ moment is $N(t) = \int_0^1 l(t,\alpha) d\alpha$.

The distribution of the parameter $\alpha$ at $t$ moment is defined as

$$P(t,\alpha) = \frac{l(t,\alpha)}{N(t)}. \qquad (1.1)$$

The population at $t$ moment is composed of defectors $D(t)$ and cooperators $C(t)$, where the numbers of defectors and cooperators are

$$D(t) \equiv \int_0^1 \alpha l(t,\alpha) d\alpha,$$

$$C(t) \equiv \int_0^1 (1-\alpha) l(t,\alpha) d\alpha = N(t) - D(t).$$

Let us denote $x(t) = D(t)/N(t)$ to be the frequency of defectors in the total population, and $y(t) = C(t)/N(t)$ to be the frequency of cooperators in the population, so that $x + y = 1$. Then the frequency of defectors

$$x(t) = \frac{D(t)}{N(t)} = \int_0^1 \frac{\alpha l(t,\alpha)}{N(t)} d\alpha = E^t[\alpha] \qquad (1.2)$$

where $E^t[\alpha]$ is the mean value of the parameter $\alpha$ at $t$ moment over the probability (1.1).

Following standard assumptions of two-strategy evolutionary game theory, we assume that cooperators have an average number of offspring per individual during the time interval $(t, t+\Delta t)$ equal to $(G_{00} y(t) + G_{01} x(t)) \Delta t$, and defectors have an average number of offspring $(G_{10} y(t) + G_{11} x(t)) \Delta t$. Consequently, we arrive at the master model

$$\frac{dl(t,\alpha)}{dt} = l(t,\alpha) \left[ (1-\alpha)(G_{00} y(t) + G_{01} x(t)) + \alpha(G_{10} y(t) + G_{11} x(t)) \right]. \qquad (1.3)$$

This equation describes the dynamics of clones in a "mixed strategy game": the outcome of this game is determined by the payoff matrix, initial distribution of the parameter $\alpha$, and the equation (1.3).

Furthermore, we may assume that each individual in the population has an initial fitness of $F_0$; a standard assumption is that $F_0$ corresponds to the logistic growth, i.e. $F_0 = r\left(1 - \dfrac{N}{K}\right)$, where $r, K$ are positive constants. Then

$$\frac{dl(t,\alpha)}{dt} = l(t,\alpha)\left[(1-\alpha)(G_{00}y(t) + G_{01}x(t)) + \alpha(G_{10}y(t) + G_{11}x(t)) + F_0\right].$$

Notice that the initial fitness of $F_0$ affects the dynamics of clone sizes and total population size but does not change the dynamics of frequencies of defectors and cooperators, or the distribution of the parameter $\alpha$. For this reason, we focus only on equation (1.3).

Taking into account that $y(t) = 1 - x(t)$, let us rewrite equation (1.3) as follows:

$$\frac{dl(t,\alpha)}{dt} = l(t,\alpha)\left[\alpha\left((G_{10} - G_{00}) + (G_{11} - G_{01} - G_{10} + G_{00})x(t)\right) + G_{00} + (G_{01} - G_{00})x(t)\right]$$

$$= l(t,\alpha)\left[\alpha((A+B)x(t) - A) + G_{00} + (G_{01} - G_{00})x(t)\right], \quad (1.4)$$

where $A = G_{00} - G_{10}$, $B = G_{11} - G_{01}$ (see (0.2)).

Denote

$$F(\alpha,t) = \alpha((A+B)x(t) - A) + G_{00} + (G_{01} - G_{00})x(t). \quad (1.5)$$

Then

$$\frac{dl(t,\alpha)}{dt} = l(t,\alpha)F(\alpha,t) \quad (1.6)$$

so $F(\alpha,t)$ is the fitness of individuals of the clone $l(t,\alpha)$.

According to the Price' covariance equation [Price 1970, Robertson 1968; see also Rice 2006] applied to (1.6),

$$\frac{dE^t[\alpha]}{dt} = Cov^t[\alpha, F(\alpha,t)] = Var^t[\alpha]((A+B)x(t) - A).$$

Taking into account that $x(t) = E^t[\alpha]$ (see (1.2), we have proven the following

**Statement 1.**
$$\frac{dx(t)}{dt} = Var^t[\alpha]((A+B)x(t) - A). \quad (1.7)$$

Due to importance of this Statement, in s.3 we provide a direct proof of (1.7).

It is clear that replicator equation (0.1) is a particular case of equation (1.7) when $Var^t[\alpha] = x(1-x)$, i.e. when the parameter $\alpha$ is distributed according to Bernoulli distribution. This means that $\alpha$ can take only two values, 0 and 1, with probabilities $P(\alpha = 1) = x$ and $P(\alpha = 0) = 1 - x$. In this case the total population is composed of only two clones, one of which is characterized by 1- strategy and has the size $l(1,t) = x(t)N(t)$ and another of which is characterized by 0-strategy and has the size $l(0,t) = (1-x(t))N(t)$.

Despite equation (0.1) being a particular case of replicator equation (1.7), there exists a key difference between these equations. Equation (0.1) can be solved explicitly using standard methods, but equation (1.7) cannot because the current variance $Var^t[\alpha]$ is unknown in general case.

One can recognize here a particular case of the well-known problem of "dynamical insufficiency" of the Price' equation: it cannot be used alone to determine dynamics of the model over time because in order to compute the mean value of a trait one needs to know its variance, and in order to compute the variance (2nd order moment), one needs to know 3rd order moment, etc. (see, e.g., Barton, Turelli, 1987; Frank, 1997). A way to overcome the dynamical insufficiency of the Price' equation was suggested in (Karev, 2010a). We use this approach in the next section.

## Solution to the replicator equation

The only way to overcome the problem of dynamical insufficiency of equation (1.7) is to compute $E^t[\alpha]$ independently; in our case we need to compute $l(t,\alpha)$ defined by equation (1.4).

In order to solve equation (1.4), we apply the HKV (hidden keystone variable) method (Karev 2012; Karev G., Kareva I., 2014). Let us introduce formally the auxiliary keystone variable by the following "escort equation":

$$\frac{dq}{dt} = x(t), q(0) = 0. \qquad (2.1)$$

Then, denoting for brevity

$$z(t) = G_{00}t - Aq(t),$$
$$w(t) = (A+B)q(t) - At, \qquad (2.2)$$

we can write the solution to equation (1.4) as

$$l(t,\alpha) = l(0,\alpha)e^{\alpha w(t) + z(t)}. \qquad (2.3)$$

Let $M_0(\delta) = \int_0^1 e^{\delta\alpha} P(0,\alpha)d\alpha$ be the mgf (moment generation function) of the initial distribution $P(0,\alpha)$ of parameter $\alpha$; we assume that this function is known. Then

$$N(t) = \int_0^1 l(t,\alpha)d\alpha = N(0)e^{z(t)} M_0(w(t)). \qquad (2.4)$$

The current distribution of the parameter $\alpha$

$$P(t,\alpha) = \frac{l(t,\alpha)}{N(t)} = P(0,\alpha)\frac{e^{\alpha w(t)}}{M_0(w(t))} \qquad (2.5)$$

and the current mean value

$$E^t(\alpha) = \int_0^1 \alpha P(t,\alpha)d\alpha = \int_0^1 \frac{\alpha e^{\alpha w(t)}}{M_0(w(t))} P(0,\alpha)d\alpha = \frac{d\ln M_0(w)}{dw}(w(t)). \qquad (2.6)$$

Taking into account equation (1.2), $x(t) = \frac{D(t)}{N(t)} = E^t(\alpha)$, we can now write equation (2.1) for the auxiliary variable $q(t)$ in a closed form

$$\frac{dq}{dt} = \frac{d}{dw}\ln M_0(w)/_{w=-At+(A+B)q(t)}, \quad q(0) = 0. \qquad (2.7)$$

Notice that due to formula (2.5), the current distribution of $\alpha$ depends only on the variable $w(t)$. According to (2.2) and (2.7),

$$\frac{dw}{dt}=-A+(A+B)\frac{dq}{dt}=-A+(A+B)\frac{d}{dw}lnM_0(w),\ w(0)=0. \qquad (2.8)$$

Equation (2.8) is the main step in our approach to solving replicator equation (1.7). With its solution, we are able to compute the distribution of mixed strategies at every time moment using an explicit formula (2.5); we can now also compute other characteristics of interest, in particular, the frequencies of cooperators and defectors by formula (2.1).

Recall that $\frac{d}{dw}lnM_0(w(t))=x(t)$ (see (2.6)); in what follows it would be useful to rewrite equation (2.8) in the form

$$\frac{dw}{dt}=-A+(A+B)x(t). \qquad (2.9)$$

In order to avoid trivial cases and technical details, we will assume as a rule, that the support of the initial distribution is an entire interval $[0,1]$, i.e. $P_0(\alpha)>0$ for all $0\leq\alpha\leq 1$. In this case the support of the current distribution $P(t,\alpha)$, according to (2.5) is also an entire interval $[0,1]$; therefore $Var^t[\alpha]>0$ for all $t$.

### Equilibria of frequencies

The main equation for dynamics of the frequency $x(t)$ in model (1.3) is (1.7). Now we are able to deduce this equation directly. To this end, let us prove that

$$\frac{dx(t)}{dt}=Var^t[\alpha]\frac{dw(t)}{dt} \qquad (3.1)$$

Indeed, according to (2.8),

$$\frac{dx(t)}{dt}=\frac{d}{dt}\frac{d}{dw}lnM_0(w(t))=\frac{d}{dw}\left(\frac{d}{dw}lnM_0(w(t))\right)\frac{dw}{dt}.$$

Next,

$$\frac{d}{dw}\left(\frac{d}{dw}lnM_0(w(t))\right)=\frac{\int_0^1\alpha^2 exp(w(t)\alpha)P(0,\alpha)d\alpha}{M_0(w(t))}-\frac{(\int_0^1\alpha exp(w(t)\alpha)P(0,\alpha)d\alpha)^2}{M_0(w(t))^2}$$

$$=E^t(\alpha^2)-(E^t(\alpha))^2=Var^t[\alpha].$$

As

$$\frac{dw}{dt} = -A + (A+B)x(t),$$

we have

$$\frac{dx}{dt} = Var^t[\alpha](-A + (A+B)x).$$

Q.E.D.

**Corollary.** Equation (3.1) shows that $\frac{dx}{dt}$ and $\frac{dw}{dt}$ are of the same sign, so $x(t)$ monotonically increases or decreases together with $w(t)$. Each equilibrium $w^*$ of equation (2.8) or (2.9) corresponds to the equilibrium value of the frequency $x^*$.

Replicator equation (0.1) allows one to find all equilibrium values of frequencies in two-strategy two-player games; the results are well known and briefly described in s.0. Equation (1.7) allows us to prove similar results for model (1.3). Firstly, let us note that equation (1.7) always has equilibria $x = 0$ and $x = 1$ because in these cases corresponding distributions are concentrated in the points $\alpha = 0$ or $\alpha = 1$ accordingly, and hence $Var[\alpha] = 0$. Stability of these equilibria can be checked via standard analysis.

Assume that the initial distribution of the parameter $\alpha$ is not concentrated in a single point. Then $Var^t[\alpha] > 0$ for all $t$, and so the existence and stability of internal equilibrium $0 < x^* < 1$ is defined by the equation $-A + (A+B)x^* = 0$. Standard analysis implies the following

**Theorem 1.**
1) the frequency $x(t)$ has an equilibrium value $0 < x^* < 1$ if and only if the values $A = G_{00} - G_{10}$ and $B = G_{11} - G_{01}$ are non-zero and of the same sign. The equilibrium $x^* = \frac{A}{A+B}$ is stable if these values are negative and it is unstable otherwise;
2) the points $x = 0$ and $x = 1$ are always equilibrium values of $x(t)$; $x = 0$ is a stable state if $A > 0$ and unstable if $A < 0$; $x = 1$ is a stable state if $B > 0$ and unstable if $B < 0$;
3) if $A = 0$ then $x(t)$ monotonically increases if $B > 0$ and monotonically decreases if $B < 0$;
4) if $B = 0$ then $x(t)$ monotonically increases if $A < 0$ and monotonically decreases if $A > 0$;
5) if $A = B = 0$; then $x(t) = const$ for all $t$ and $P(t, \alpha) = P_0(\alpha)$ for all $t$.

Overall, all equilibria of general equation (1.7) are identical to corresponding equilibria of standard equation (0.1), but dynamics of the frequency $x(t)$ may be different. An example is given below (see s.5, Figure 7).

Now we are able to study not only dynamics of frequency $x(t) = E^t(\alpha)$, but also the dynamics of distribution of the parameter $\alpha$. For any specific game, this dynamics is completely determined by the values $A$, $B$, and the mgf $M_0$ of the initial distribution of parameter $\alpha$. Let us now summarize the main formulas, which will be used later:

$$P(t,\alpha) = P(0,\alpha) \frac{e^{\alpha w(t)}}{M_0(w(t))}, \qquad (3.2)$$

$$x(t) = \frac{d\ln M_0(w)}{dw}(w(t)) \qquad (3.3)$$

where $w(t)$ is the solution to the Cauchy problem

$$\frac{dw}{dt} = -A + (A+B)\frac{d}{dw}\ln M_0(w(t)), \quad w(0) = 0. \qquad (3.4)$$

### Dynamics of the distribution of strategies

*Replicator equation*

The process of natural selection of strategies in model (1.3) is captured by the evolution of distribution of the parameter $\alpha$, $P(t,\alpha) = \frac{l(t,\alpha)}{N(t)}$, where $N(t) = \int_A l(t,\alpha) d\alpha$ and

$\frac{dl(t,\alpha)}{dt} = l(t,\alpha) F(t,\alpha)$ (see (1.6)).

It is well known (see, e.g., [Hofbauer, Zigmund, 1998]) that the pdf $P(t,a)$ solves the replicator equation of the form

$$\frac{dP(t,\alpha)}{dt} = P(t,\alpha)\bigl(F(t,\alpha) - E^t[F(t,\alpha)]\bigr).$$

According to definition (1.5) of the function $F(t,\alpha)$,

$$F(t,\alpha) - E^t[F(t,\alpha)] = ((A+B)x(t) - A)(\alpha - E^t[\alpha]).$$

Hence, as $x(t) = E^t[\alpha]$ (see (1.2)),

$$\frac{dP(t,\alpha)}{dt} = P(t,\alpha)((A+B)x(t) - A)(\alpha - x(t)). \tag{4.1}$$

The solution to this equation is given by formulas (3.2) - (3.4).

Now, in order to investigate the process of natural selection of strategies, we may take initial distribution of the parameter as uniform in $[0,1]$; in this case

$$M_0(\delta) = \int_0^1 e^{\delta\alpha} d\alpha = \frac{(e^\delta - 1)}{\delta}. \tag{4.3}$$

A more general case is the truncated exponential distribution in $[0,1]$ of the form

$$P(0,\alpha) = \frac{s}{1-e^{-s}} e^{-s\alpha}, 0 \leq \alpha \leq 1 \tag{4.4}$$

with the mean value $E[\alpha] = \frac{1}{s} + \frac{1}{1-e^s}$ and mgf

$$M_0(\delta) = \frac{s}{1-e^{-s}} \frac{e^{\delta-s} - 1}{\delta - s}. \tag{4.5}$$

Then

$$\frac{d}{dw} \ln M_0(w) = \frac{1}{1-e^{s-w}} + \frac{1}{s-w}. \tag{4.6}$$

The uniform distribution is a limit case of truncated exponential as $s \to 0$.

One more representative case is normal distribution truncated on the interval $[0,1]$. The pdf of this distribution depends on two parameters, $m$ and $s$, and has a form

$$P(0,\alpha) = C e^{-\frac{(\alpha-m)^2}{s}}, \ 0 \leq \alpha \leq 1 \tag{4.7}$$

where normalization constant $C = 2 / \left[ \sqrt{\pi s} \left( Erf\left(\frac{1-m}{\sqrt{s}}\right) + Erf\left(\frac{m}{\sqrt{s}}\right) \right) \right]$.

The mgf of this distribution is given by the formula

$$M_0(w) = \frac{e^{mw+\frac{sw^2}{4}}\left(-Erf\left[\frac{-2+2m+sw}{2\sqrt{s}}\right]+Erf\left[\frac{2m+sw}{2\sqrt{s}}\right]\right)}{Erf\left[\frac{1-m}{\sqrt{s}}\right]+Erf\left[\frac{m}{\sqrt{s}}\right]} \quad (4.8)$$

In the following sections we investigate the dynamics of parameter $\alpha$ that describes the process of natural selection of a mixed strategy in different games. A good introduction and description of the considered games can be found, e.g., in (Broom& Rychtár, 2013, ch.4; Gintis, 2009, ch.3).

### Prisoner's dilemma

A large body of literature is devoted to the discussion and investigation of different versions of the game known as Prisoner's dilemma (PD). In this game, defection strategy gives the player a higher payoff than cooperation strategy irrespective of which strategy is used by the second player; therefore, a rational individual should defect. However, both players would get higher payoff if they cooperated, which makes this a dilemma (the mathematical example and corresponding payoff matrix is shown below).

The PD-game is widely applied to many social and behavioral problems, but also has (perhaps, limited) applications to some biological situations, see, e.g., (Broom& Rychtár, 2013, ch.4, Turner & Lin Chao, 1999); an extreme case of the PD-game is the well-known "tragedy of the commons", a situation when the strategy of over-exploitation of a shared resource on which the population depends (defecting) rather than preserving it (cooperating) can lead to resource destruction and consequent population collapse (Hardin 1998).

It is well known that the frequency of defectors tends to 1 for a two-person two strategy PD- game. This statement is also true for mixed-strategy PD game. Elements of the payoff matrix of such a PD game satisfy the inequalities $G_{10} > G_{00} > G_{11} > G_{01}$, so that $A = G_{00} - G_{10} < 0$, $B = G_{11} - G_{01} > 0$. Then, according to assertions 1) and 2) of Theorem 1, $\lim_{t\to\infty} x(t) = 1$, and therefore the final distribution of the parameter $\alpha$ is concentrated in the point $\alpha = 1$ independently of initial distribution. Additionally, we can trace the process of strategy selection described by the evolution of the parameter $\alpha$ distribution.

A standard example of PD game is defined by the payoff matrix
$$G_{PD} = \begin{pmatrix} b-c & -c \\ b & 0 \end{pmatrix}, \ b > c > 0.$$

Then $A = -c, B = c$ and $\dfrac{dw}{dt} = c$ (see equation (3.4)), hence

$w(t) = ct$ and $P(t,\alpha) = P(0,\alpha) \dfrac{e^{\alpha ct}}{M_0(ct)}$.

We can see that the dynamics of the distribution of parameter $\alpha$ is identical to the dynamics of the distribution of the Malthusian parameter in the simplest inhomogeneous Malthusian model given by the equation

$$\frac{dl(t,\alpha)}{dt} = \alpha c l(t,\alpha).$$

The solution to this equation is $l(t,\alpha) = l(0,\alpha) e^{\alpha ct}$, then

$$P(t,\alpha) = l(t,\alpha) / \int_0^1 l(t,\alpha) d\alpha = \frac{P(0,\alpha) e^{\alpha ct}}{M_0(ct)}.$$

This model describes dynamics of a population, which consists of clones $l(t,\alpha)$ that grow exponentially and independently of each other; the model was studied in detail in application to global demography in [Karev 2005].

Consider a numerical example of mixed strategy PD game with the payoff matrix

$G_{PD} = \begin{pmatrix} 1 & -1 \\ 2 & 0 \end{pmatrix}$; then $A = -1, B = 1, \ w = t, \ P(t,\alpha) = \dfrac{P(0,\alpha) e^{\alpha t}}{M_0(t)}$.

Dynamics of the distribution of parameter $\alpha$ is shown in Figure 1.

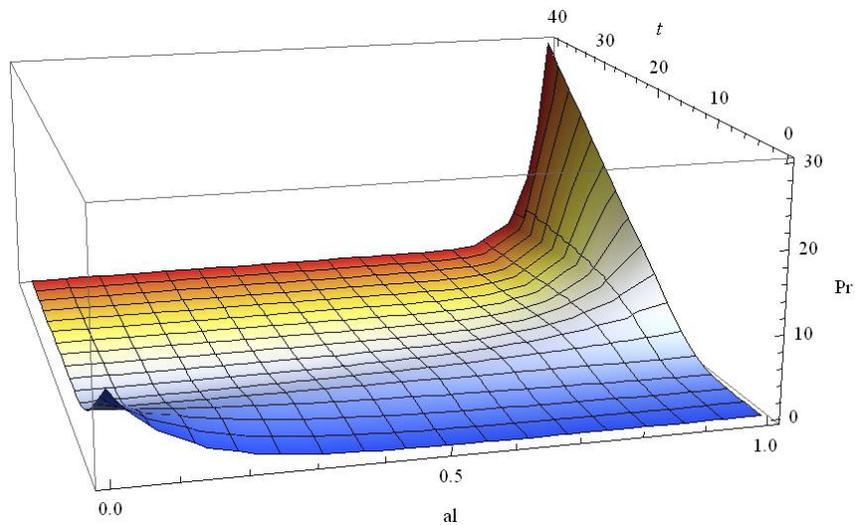

**Figure 1**. Dynamics of the distribution of parameter $\alpha$. Initial distribution (blue) is truncated exponential (Equation 4.4) with $s = 10$; the final distribution (red) is concentrated in the point $\alpha = 1$.

The following Figure 2 shows the dynamics of initial normal truncated distribution of the parameter $\alpha$.

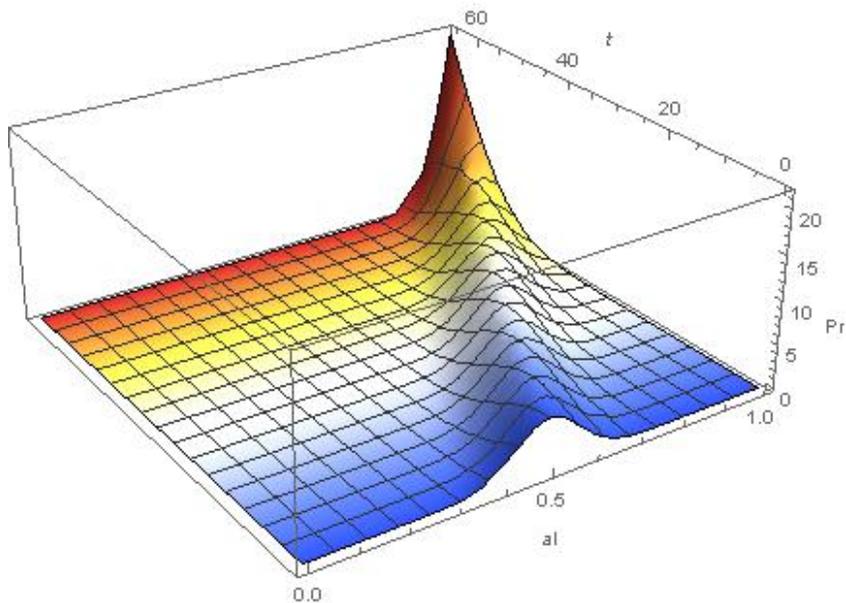

**Figure 2.** Dynamics of the initial (blue) truncated normal distribution (Equation 4.7) with $m = 0.5, s = 0.01$ to the final distribution (red), concentrated in the point $\alpha = 1$.

Coordination Game or SH game.

The Coordination game, also known as Stag Hunt (SH) game, is characterized by bi-stability, when both equilibria $x=0$ and $x=1$ are stable and there exists an unstable equilibrium $0 < x^* < 1$ that divides the domains of attraction of the stable equilibria. This game is defined by the payoff matrix, whose elements satisfy $G_{00} > G_{10}, G_{11} > G_{01}$.

In this case

$A = G_{00} - G_{10} > 0, B = G_{11} - G_{01} > 0$, and $x^* = \frac{A}{A+B}$.

Evolution of the distribution of parameter $\alpha$ crucially depends on the initial mean value $E^0[\alpha]$; if $E^0[\alpha] < x^*$, then the distribution is concentrated in course of time in the point $\alpha = 0$, and if $E^0[\alpha] > x^*$, then the distribution is concentrated in course of time in the point $\alpha = 1$.

Consider the following numerical example. Let $A=B=1$, then $x^* = \frac{1}{2}$. Now assume the initial distribution is truncated exponential. If $s = 0.01$ then $E^0[\alpha] = 0.4992$, and the distribution of $\alpha$ is concentrated in the point $\alpha = 0$; if $s = -0.01$, then $E^0[\alpha] = 0.5008$, and the distribution of $\alpha$ is concentrated in the point $\alpha = 1$ (see Figure 3). As we can see, in this game a very small difference in the initial distribution can lead to dramatic differences in the outcomes of natural selection of strategies.

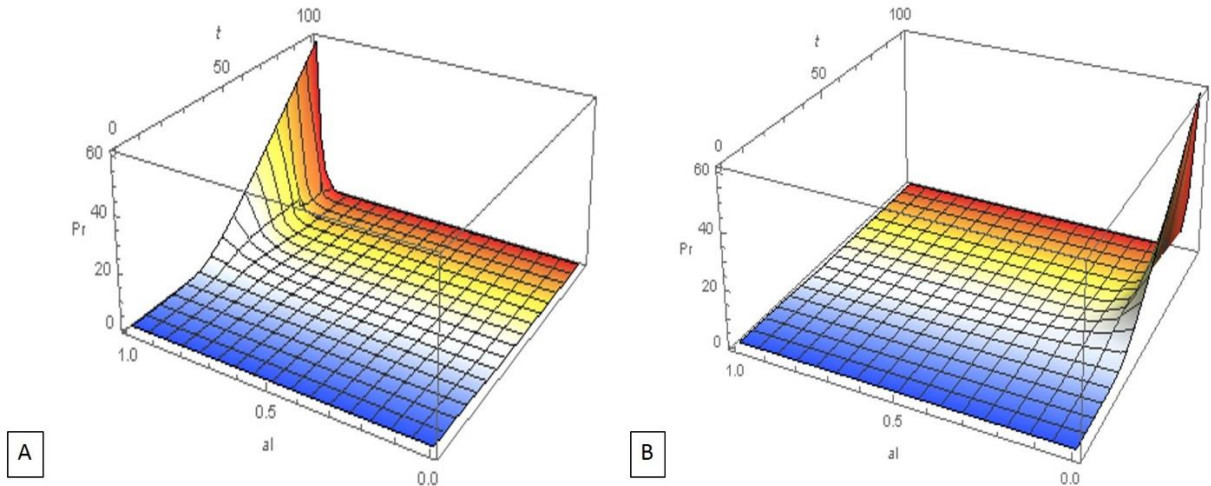

**Figure 3**. Evolution of initial truncated exponential distribution (Equation 4.4). Here the initial conditions are (A) $E^0[\alpha] = 0.4992$ (left panel) and (B) $E^0[\alpha] = 0.5008$ (right panel).

Now let the initial distribution be the truncated normal. Let us once again consider the SH game with $A=B=1$. Dynamics of the initial distribution is shown on Figure 4.

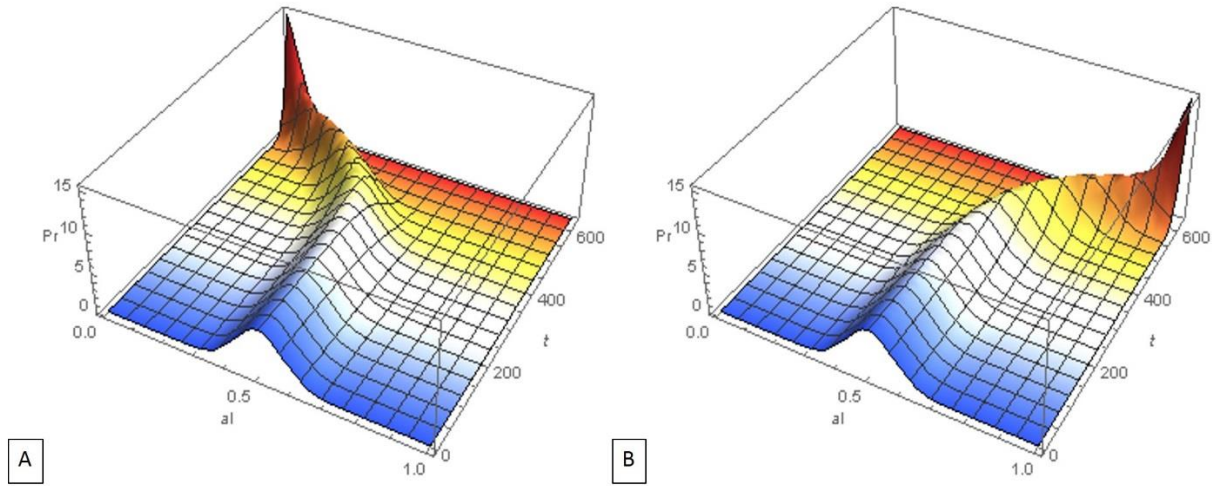

**Figure 4**. Evolution of initial truncated normal distribution (Equation 4.7) with $m = 0.5$. Initial conditions are $E^0[\alpha]$=0.499 (left panel) and $E^0[\alpha]$=0.501 (right panel).

The SH-game is an example of situation when the selection of one or another pure strategy crucially depends on the initial composition of the population, and so any random perturbation can have dramatic effects of the predicted outcome.

## Natural selection of strategies in a "Hawk- Dove" game

The "Hawk- Dove" (HD-) game was introduced by Maynard Smith (1974) and can be described as follows. Consider individuals within a population competing for resources V. Doves avoid confrontation, while hawks provoke fights. Thus, on average, two doves will share the resources and the average increase in fitness for doves is V/2. A dove that encounters a hawk will leave all the resource to the hawk. If a hawk meets a hawk, they escalate until one of the two gets knocked out. The winner's fitness is increased by V, while loser's fitness is reduced by C, so that the average increase in fitness is (V - C)/2, which is negative if the cost of the injury exceeds the gain from winning the fight.

The importance of HD game was clearly explained by [Webb 2007, s.8.4]: "The biological significance of the Hawk-Dove game is that it provides an alternative to group-selectionist arguments for the persistence of species whose members have potentially lethal attributes (teeth, horns, etc.). The question to be answered is the following. Because it is obviously advantageous to fight for a resource (having it all is better than sharing), why don't animals always end up killing (or at least seriously maiming) each other? The group-selectionist answer is that any species following this strategy would die out pretty quickly, so animals hold back from all out contests "for the good of the species". The "problem" with this is that it seems to require more than just individual-based Natural Selection to be driving Evolution. So, if group selection is the only possible answer, then that would be a very important result. However, the Hawk-Dove game shows that there is an alternative – one that is based fairly and squarely on the action of Natural Selection on individuals. So, applying Occam's Razor, there is no need to invoke group selection."

Qualitative behavior of strategies in this game is characterized by the existence of a stable polymorphic state $0 < x^* < 1$. This state exists if the following inequalities hold: $G_{00} < G_{10}$ and $G_{11} < G_{01}$; then $A = G_{00} - G_{10} < 0$ and $B = G_{11} - G_{01} < 0$. It follows from Statement 1 that in this case $x^* = \dfrac{A}{A+B}$ is an equilibrium frequency of defectors (hawks).

This result is well known for the two-strategy HD game. However, the approach developed in previous sections allows us to expand it by tracing the process of natural selection of strategies for a mixed strategy HD game to obtain some new and interesting results. Notice that in cases of PD and SH mixed strategy games the final distribution of strategies is always concentrated in the points $\alpha = 0$ or $\alpha = 1$, i.e. only one pure strategy can be selected over time. In contrast, the set of possible limit stationary distributions for HD game is extremely rich. The distribution $P(t,\alpha)$ is stationary, by definition, if $\dfrac{dP(t,\alpha)}{dt} = 0$ for all $\alpha$. The dynamics of current distribution $P(t,\alpha)$ is defined by the replicator equation (4.1),

$$\frac{dP(t,\alpha)}{dt} = P(t,\alpha)\big((A+B)x(t) - A\big)\big(\alpha - x(t)\big).$$

Hence, if the frequency of defectors is given by $x = \dfrac{A}{A+B}$, then the distribution $P(t,\alpha)$ is stationary. Recall that $x(t) = E^t[\alpha]$ (see equation (1.2)). We have proven the following

**Statement 2.** Let $0 < \dfrac{A}{A+B} < 1$ and $P(\alpha)$ be any distribution of $\alpha \in [0,1]$ such that $E[\alpha] = \dfrac{A}{A+B}$. Then $P(\alpha)$ is a stationary distribution.

The current distribution of the parameter $\alpha$ given the initial distribution $P(0,\alpha)$ is described by the equations (3.2). Then, if $0 < \dfrac{A}{A+B} < 1$, a non-singular limit stationary distribution of $\alpha$ exists and has the form

$$P^*(\alpha) = \dfrac{P(0,\alpha)e^{\alpha w^*}}{M_0(w^*)} \tag{5.1}$$

where $w^*$ solves the equation $\dfrac{dw}{dt} = 0$, i.e. (see (3.4))

$$\dfrac{d}{dw}\ln M_0(w^*) = E^*[\alpha] = \dfrac{A}{A+B}.$$

Here we denote $E^*[\alpha]$ to be the mean value of $\alpha$ with respect to distribution $P^*(\alpha)$. It follows from (5.1) that all mixed strategies that were present in the population initially do not disappear over time but remain in the population indefinitely.

A classic example of a Hawk-Dove game is defined by the following payoff matrix:

$$G = \begin{pmatrix} & Coop & Def \\ Coop & V/2 & 0 \\ Def & V & \dfrac{V-C}{2} \end{pmatrix}, \; C > V. \tag{5.2}$$

Here $A = -\dfrac{V}{2}, B = \dfrac{V-C}{2} < 0, A + B = -\dfrac{C}{2}$. Consequently, equation (1.7) becomes

$$\frac{dx(t)}{dt} = Var^t[\alpha]((A+B)x(t) - A) = Var^t[\alpha]\left(\frac{V}{2} - \frac{C}{2}x(t)\right) \qquad (5.3)$$

and the equilibrium frequency of defectors is $x^* = \dfrac{V}{C}$.

In addition to these known results, equations (2.5) and (2.8) allow us to study the dynamics of distribution of the parameter $\alpha$ that describes the process of natural selection of strategies. Equations (2.8) now becomes

$$\frac{dw}{dt} = \frac{V}{2} - \frac{C}{2}\frac{d}{dw} lnM_0(w(t))\,.$$

This equation has a unique stable equilibrium $w^*$, which solves the equation

$$\frac{d}{dw} lnM_0(w) = \frac{V}{C}\,.$$

If the initial distribution of $\alpha$ is truncated exponential (4.4). then

$$\frac{d}{dw} lnM_0(w) = \frac{1}{1-e^{s-w}} + \frac{1}{s-w},$$

so

$$\frac{dw}{dt} = \frac{V}{2} - \frac{C}{2}\left(\frac{1}{1-e^{s-w}} + \frac{1}{s-w}\right) \qquad (5.4)$$

and $w^*$ solves the equation $\dfrac{1}{1-e^{s-w^*}} + \dfrac{1}{s-w^*} = \dfrac{V}{C}$.

Now that we have a solution to equation (5.4), we can compute the pdf $P(t,\alpha)$ according to formula (2.6) for all $t$, and therefore can describe all the statistical characteristics of this game. Consider the following numerical example, where $V = 2, C = 6$. The dynamics of the frequency $x(t)$ of defectors depends on initial distribution of $\alpha$, and in particular on the parameter $s$ of the truncated exponential distribution (4.4). However, the final value $x^*$ does not depend on the initial distribution, as can be seen in Figure 5.

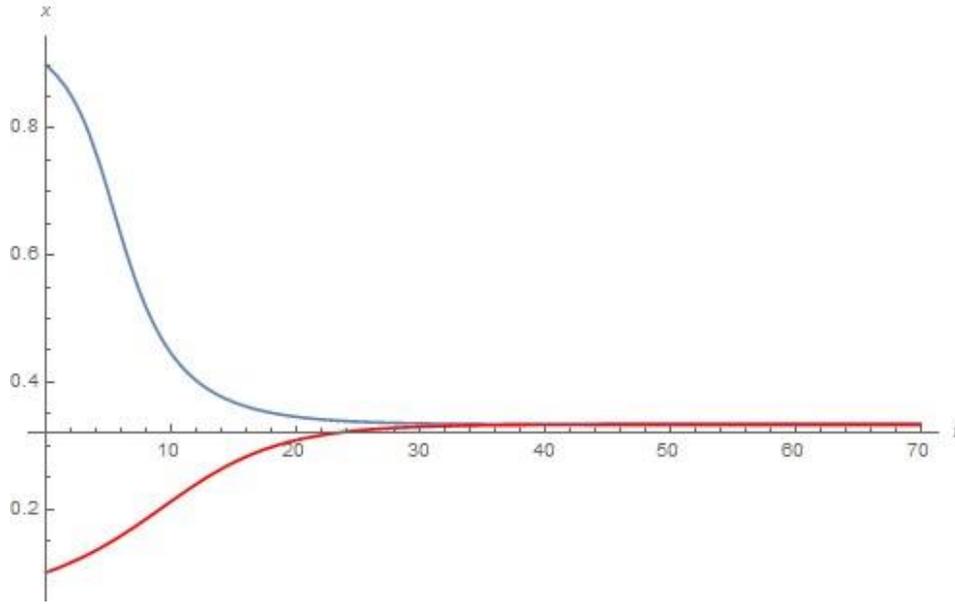

**Figure 5**. Dynamics of the defector frequency in the Hawk-Dove game as described by Equation (5.6). The initial distribution of $\alpha$ is truncated exponential with parameter *s*, as defined in Equation (4.4); s=10 (red), s= -10 (blue); the limit frequency of defectors $x^* = \dfrac{V}{C} = \dfrac{1}{3}$.

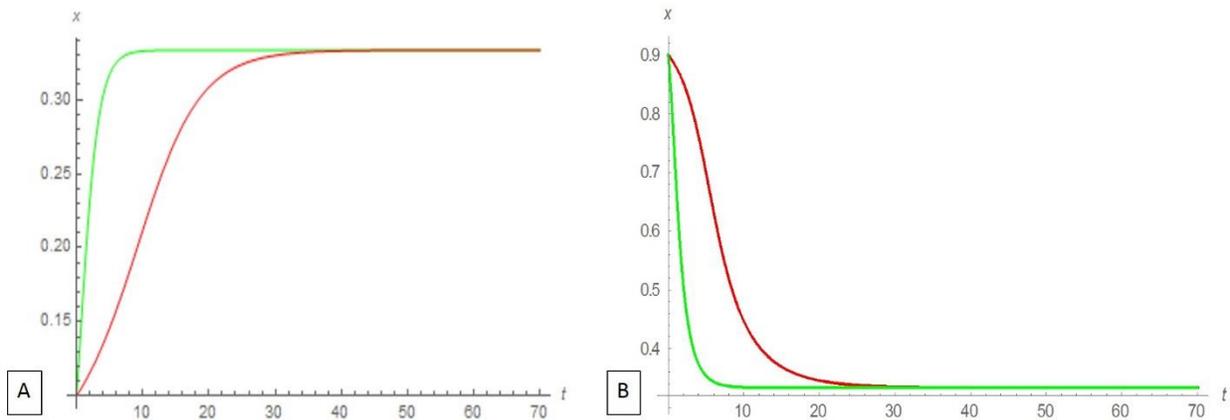

**Figure 6**. The Hawk-Dove game. Comparison of the dynamics of the frequency of defectors defined by replicator equation (0.1) (green) and by equation (5.3) (red), with $A = -1, B = -2$. The initial distribution of $\alpha$ is truncated exponential in [0,1] with parameter *s*, as defined in Equation (4.4). (A) Left panel: $x(0) = E^0[\alpha] = 0.1, s = 10$; (B) right panel: $x(0) = E^0[\alpha] = 0.9, s = -10$.

Complete information about the process of natural selection of strategies can be obtained from dynamics of the distribution of parameter $\alpha$. Let the initial distribution be truncated exponential. Then the current distribution at the moment $t$ is also truncated exponential with pdf

$$P(t,\alpha) = \frac{e^{\alpha(w(t)-s)}(-w(t)+s)}{1-e^{w(t)-s}}.  \tag{5.5}$$

The limit distribution is

$$P^*(\alpha) = \frac{e^{\alpha(w^*-s)}(-w^*+s)}{1-e^{w^*-s}},$$

where $w^*$ is the root of the equation

$$\frac{e^s + e^w(-1-s+w)}{(e^s - e^w)(s-w)} = V/C. \tag{5.6}$$

The dynamics of distribution of the parameter $\alpha$ given by formula (5.5) is shown in Fig. 7.

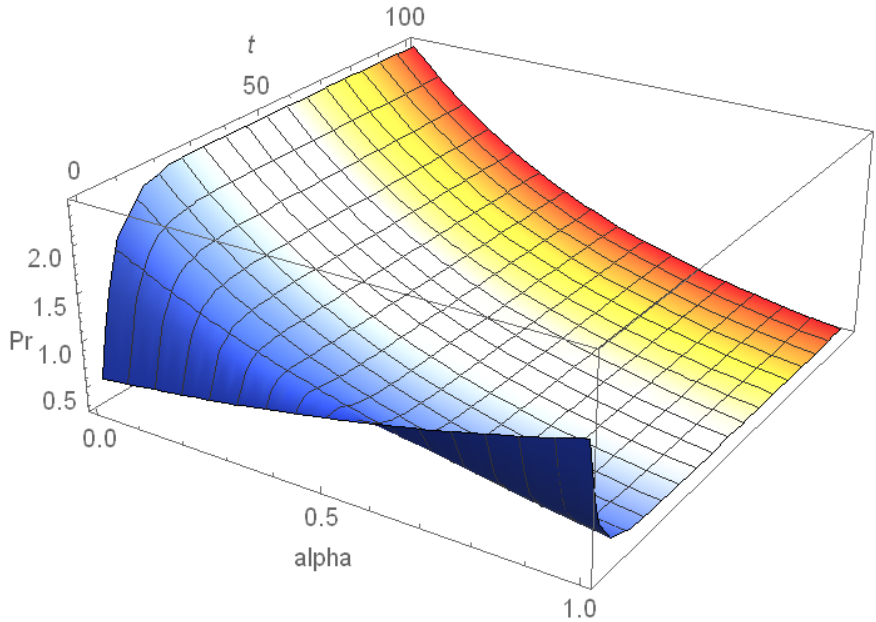

**Figure 7.** The dynamics of initial truncated exponential distribution of $\alpha$ in the Hawk-Dove game with s=-1 (blue).

We have seen that the total frequency of defectors, which is equal to $E^t[\alpha]$, tends to $\frac{V}{C} = 1/3$ in accordance with known results. A new piece of information is that the distribution of mixed strategies characterized by the parameter $\alpha$ tends over time to the truncated exponential distribution, whose mean value is equal to the limit frequency of defectors.

Notice that the support of all current and limit distributions coincides with the support of initial distribution regardless of what initial distribution was. This means that all mixed strategies that are present initially will also be present in the final state of the system.

Let us now assume that the initial distribution is normal truncated in the interval [0,1]. This distribution depends on two parameters, $m$ and $s$, and has pdf defined by equation (4.7) and mgf defined by equation (4.8).

The current pdf is given by the formula

$$P(t,\alpha) = \frac{2e^{-\frac{(-\alpha + m + sw/2)^2}{s}}}{\sqrt{\pi}\sqrt{s}\left(-\text{Erf}\left[\frac{-1+m+sw/2}{\sqrt{s}}\right] + \text{Erf}\left[\frac{m+sw/2}{\sqrt{s}}\right]\right)}$$

where $w(t)$ is the solution to the equation

$$\frac{dw}{dt} = -A + (A+B)\left(m + sw/2 + \frac{e^{-\frac{1+(m+sw/2)^2}{s}}\left(e^{\frac{1}{s}} - e^{\frac{2m}{s}+w}\right)\sqrt{s}}{\sqrt{\pi}\left(-\text{Erf}\left[\frac{-1+m+sw/2}{\sqrt{s}}\right] + \text{Erf}\left[\frac{m+sw/2}{\sqrt{s}}\right]\right)}\right).$$

A typical evolution of truncated normal distribution is shown on Figure 8.

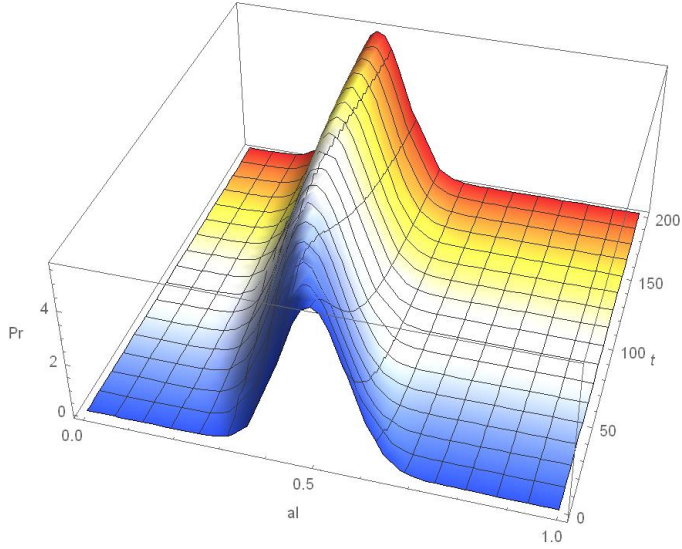

**Figure 8.** Evolution of initial truncated normal distribution (blue) in the Hawk-Dove game with $A=-1, B=-2$. Parameters of initial distribution are $m=\frac{1}{2}, s=0.01$; the final distribution (red) is truncated normal with $m=\frac{1}{3}, s=0.01$.

The "game of chicken" is just a version of the HD- game and differs only in interpretation. A standard example of this game is given by the following payoff matrix:

$$G = \begin{pmatrix} b-c & b-2c \\ b & 0 \end{pmatrix}, \ b>0, c>0, \ b-2c>0.$$

For this game $A=-c$, $B=2c-b<0$, so condition 1) of Theorem 1 holds and

$0 < x^* = \frac{A}{A+B} = \frac{c}{b-c} < 1$ is a stable limit frequency of defectors.

The dynamics of defector frequency is defined by the equation (1.7), which now reads

$$\frac{dx(t)}{dt} = Var^t[\alpha](c+(c-b)x(t)).$$

If $2c > b > c$, then $B > 0$, and the 'game of chicken' becomes Prisoner's Dilemma, where $x^* = 1$ is a single equilibrium frequency of defectors. If $2c = b$ then $x^* = 1$ is also a single limit value of the frequency.

Figure 9 illustrates the dynamics of frequency of defectors for the 'game of chicken', and its transition to Prisoner's Dilemma when the value of $b$ decreases. Blue and green curves correspond to the case $b > 2c$ (game of chicken), and red curve corresponds to the case $b < 2c$ (PD game).

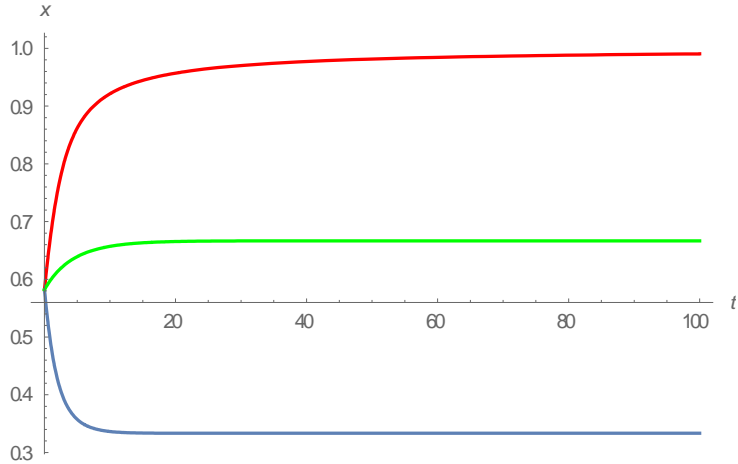

**Figure 9**. Dynamics of defector frequency for the 'game of chicken'. The initial distribution of $\alpha$ is truncated exponential in [0,1] with parameter $s = -1$, as defined in Equation (4.4). Parameters for the blue curve are $b = 8, c = 2$, and for the green curve are $b = 5, c = 2$. With change of parameter $b$, the 'game of chicken' becomes the PD-game, for red curve being $b = 3, c = 2$.

We are also able to trace the dynamics of the distribution of $\alpha$ given by formulas (3.2), (3.4) where now

$$\frac{dw}{dt} = c + (c-b)\frac{d}{dw}\ln M_0(w) .$$

If the initial distribution of $\alpha$ is truncated exponential (4.4) then this last equation becomes

$$\frac{dw}{dt} = c + (c-b)\frac{e^s + e^w(-1-s+w)}{(e^s - e^w)(s-w)}. \tag{5.7}$$

Dynamics of the current distribution of $\alpha$ is shown in Figures 5 and 6. If $b = 8, c = 2$, then we have HD-game (or 'game of chicken') and the final distribution is non-singular (see Fig. 10, left panel) with $0 < E[\alpha] < 1$. If however $b = 3, c = 2$, then we have the PD game, and the final distribution is concentrated in the point $\alpha = 1$, see Fig.10, right panel.

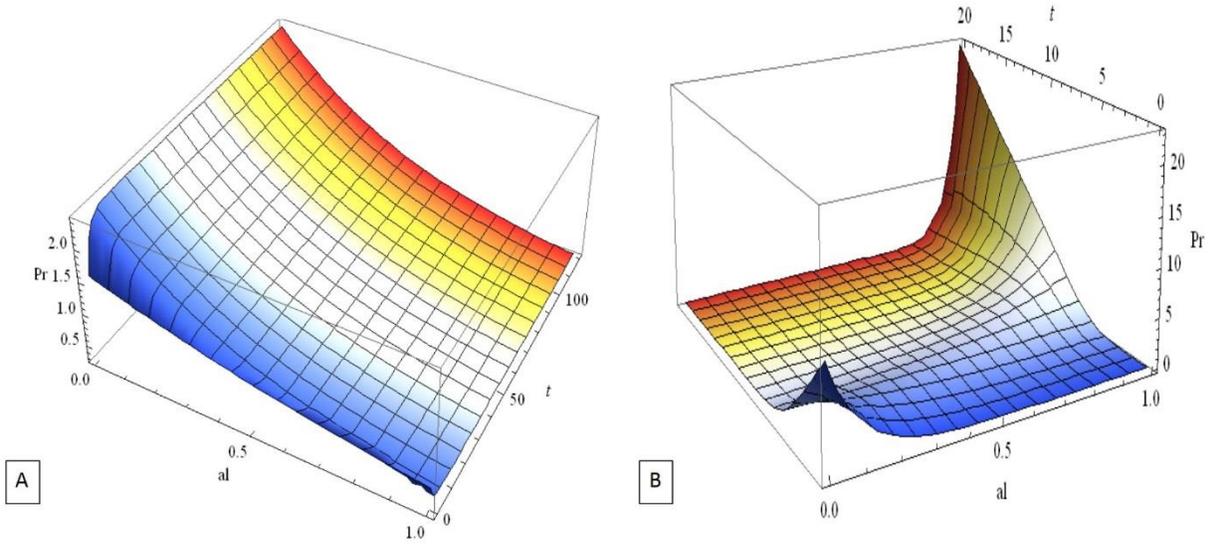

**Figure 10.** Dynamics of initial truncated exponential distribution of $\alpha$, with parameter s=1; the initial distribution is shown in blue, the final distribution is shown in red. (A) Left panel with $b=8, c=2$, depicts the HD-game with $A=-2, B=-6$; the final distribution is truncated exponential. (B) Right panel with $b=3, c=2$ depicts the PD-game with $A=-2, B=1$, the final distribution is concentrated in the point $\alpha=1$.

## Natural selection of strategies and the Principle of minimum of information gain

In this section we show that the dynamics of the distribution of strategies as it was described by formulas (3.2) - (3.4) obey the Principle of Minimum Information Gain for all considered games.

The Principle was developed in information statistics and is based on the hypothesis that given a prior probability distribution *m* and precisely defined prior data, the probability distribution *p* that best represents the current state of knowledge is the distribution that minimizes KL-divergence between *p* and *m* (after Kullback-Liebler, 1951). The KL-divergence $I[p:m]$ is a measure of information gain in moving from a prior distribution *m* to a posterior distribution *p*; it is defined by the formula

$$I[p:m] = \int_A p(\alpha)\ln\frac{p(\alpha)}{m(\alpha)}d\alpha = E_p[\ln\frac{p}{m}]. \qquad (6.1)$$

$I[p:m]$ is known also as relative or cross-entropy in natural sciences; therefore, the principle of minimum of information gain is mathematically equivalent to the Principle of Minimum of relative or cross- entropy, known as MinxEnt for brevity.

The MinxEnt was successfully applied to various statistical, physical and biological problems as a method for inference of unknown distributions, subject to some given constraints. The importance of the Principle of minimum of relative entropy and its special case, the Principle of maximum entropy, was demonstrated by Jaynes and his followers, who showed that essentially all known statistical mechanics can be derived from this principle; see (Jaynes 1957, 2003) for further details and applications of the MaxEnt principle.

A serious objection against this approach in natural sciences is that the principle of maximum of entropy does not follow from the basic laws and hence may not be postulated as an independent assertion but should follow from the system dynamics. The problem was clearly formulated by A. Einstein (1910): "…the statistics of a system should follow from its dynamics and, in principle, could not be postulated a priori". It is a way to obtain a satisfactory rationale of the MaxEnt principle, but it was not realized (and probably could not be realized) in statistical physics.

It was shown in (Karev 2010b) that within the frameworks of mathematical theory of selection, MinxEnt is not an external independent hypothesis but is a strong mathematical assertion, which follows from the dynamics of corresponding models. Below we give a direct proof of this statement in application to our models of strategy selection.

A general theory for applications of the Principle of minimum of information gain was developed in (Kullback 1997). The probability density function (pdf) $m(\alpha), \alpha \in A$ in (6.1) is assumed to be given. It is also assumed that posterior data in the form of expected value of some variable, $T(\alpha)$, is given, $E^p[T] = C = const$. Then the distribution that provides minimum of $I[p:m]$ subject to the constraint $E^p[T] = C$ is given by the formula (Kullback 1997, Theorem 2.1):

$$p(\alpha) = m(\alpha)\frac{e^{wT(\alpha)}}{M(w)}. \qquad (6.2)$$

Here $M(w) = \int_A e^{wT(\alpha)} m(\alpha) d\alpha$, and the multiplier $w$ is the solution to the equation

$$\frac{d}{dw} \log(M(w)) = C. \tag{6.3}$$

If pdf $p$ is defined this way, then

$$I[p:m] = Cw - \log(M(w)). \tag{6.4}$$

For example, if there is no any additional information, then the MinxEnt distribution (6.2) in an interval is uniform; if the mean of the distribution is the only estimated quantity, then the MinxEnt distribution (for $\alpha \geq 0$) is exponential with the estimated mean; if the mean and variance are estimated, then the MinxEnt distribution (for $-\infty < \alpha < \infty$) is normal with the estimated mean and variance; see (Kapur, 1989) for these and other examples.

Now let us apply Kullback's theorem to our problem of strategy selection using "inverse logic". That is, we do not seek an unknown distribution $P_t$ that would minimize the KL-divergence $I[P_t : P_0]$ subject to the mean value $E^t[\alpha] = x(t)$. Instead, we already know the distribution of strategies given by formula (2.5) and the mean value $E^t[\alpha]$ given by equation (2.6) at each time moment. Let us plug into formulas (6.1) - (6.4) the following quantities: $A = [0,1], T(\alpha) = \alpha$, $m = P_0$, $p = P_t$. Then it can be easily seen that distribution (2.5) coincides with the distribution (6.2) that minimizes the information gain $I[P_t : P_0]$ subject to the given mean value $E^t[\alpha] = x(t)$. Therefore, the following theorem holds:

**Theorem 2.**

*The current strategy distribution* (2.5) *provides minimum of information gain* $I[P_t : P_0]$ *over all probability distributions* $P_t = P(t, \alpha)$ *with a given value of* $E^t(\alpha)$ *at every time moment t;*

*the value* $I[P_t : P_0]$ *can be computed using the formula*

$$I[P_t : P_0] = w(t)x(t) - \ln M_0(w(t)) = w(t)\frac{d\ln M_0(w(t))}{dw} - \ln M_0(w(t)). \tag{6.5}$$

Let us emphasize that Theorem 2 is valid for any mixed strategy game described by equation (1.3).

Theorem 2 implies an interesting

**Corollary**.

$$I[P_t : P_0] = \ln P(t, x(t)) - \ln(P(0, x(t))). \tag{6.6}$$

Indeed, according to formulas (6.5) and (3.2),

$$\exp(I[P_t : P_0]) = \frac{e^{w(t)x(t)}}{M_0(w(t))} = \frac{P(t, x(t))}{P(0, x(t))}. \tag{6.7}$$

**Statement 3**. Let the pdf of initial distribution $0 < P(0, \alpha) < \infty$ for all $\alpha \in [0,1]$. Then information gain $I[P_t : P_0]$ tends to a *finite* value for HD-game and $I[P_t : P_0] \to \infty$ for PD, H and SH games as $t \to \infty$.

Indeed, it was proven in Theorem 1 that $\lim_{t \to \infty} x(t) = x^*$ as $t \to \infty$, $0 < x^* = \frac{A}{A+B} < 1$, $A < 0$, $B < 0$ for HD-game. Then equation (2.9) has a finite stable equilibrium $w^*$. Hence, $w(t) \to w^*$ as $t \to \infty$, and so $I[P_t : P_0] \to w^* x^* - \ln M_0(w^*)$.

The second assertion of the statement follows from formula (6.6). Let, for example, $\lim_{t \to \infty} x(t) = 1$ as in PD-game. Then the distribution of the parameter $\alpha$ over time becomes concentrated in the point $\alpha = 1$. This means that the density of this distribution, $P(t, \alpha) \to \delta(1)$, where $\delta(\alpha)$ is the Dirac $\delta$-function. Then $P(t, x(t)) \to \infty$ as $t \to \infty$. Therefore $\frac{P(t, x(t))}{P(0, x(t))} \to \infty$ as $\lim_{t \to \infty} P(0, x(t)) = P(0,1) < \infty$. Then it follows from (6.7) that $\lim_{t \to \infty} I[P_t : P_0] = \infty$.

The case $\lim x(t) \to 0$ can be considered in same way. Q.E.D.

The following figures show the dynamics of KL-divergence $I[P_t : P_0]$ between initial and current strategy distributions for different games considered in ss. 3-5. In all cases the initial distribution was taken as truncated exponential according to equation (4.4) with parameter *s*.

Figure 11 shows $I[P_t : P_0]$ for Prisoner's Dilemma and Harmony games. Recall that for these games the distribution $P(t,\alpha)$ becomes concentrated over time in the points $\alpha = 1$ and $\alpha = 0$ accordingly, independently of the initial distribution. In both cases, $I[P_t : P_0] \to \infty$ as $t \to \infty$.

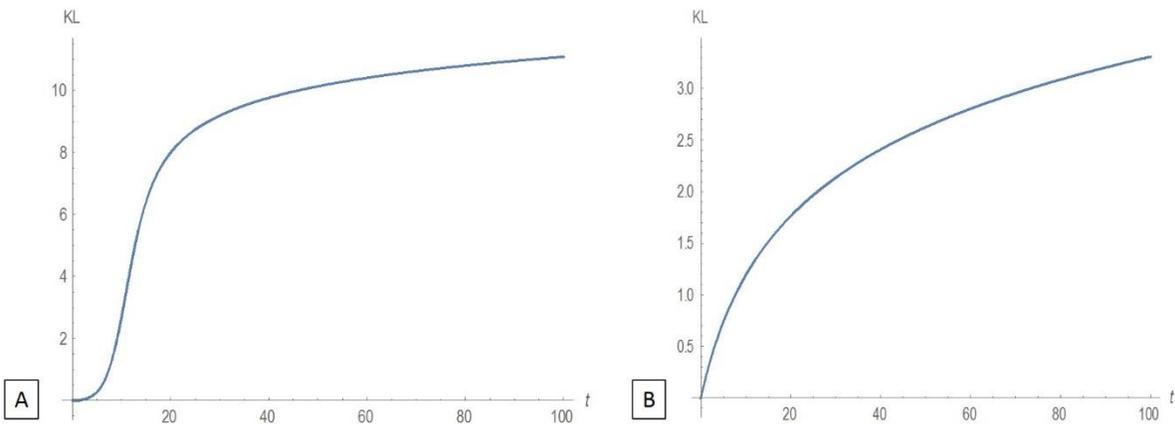

**Figure 11**. Dynamics of KL-divergence as defined in Equation (6.1). (A) Left panel: Prisoner's Dilemma, $A = -1, B = 1, s = 10$. (B) Right panel: Harmony game, $A = 1, B = -1, s = 10$.

Figure 12 shows dynamics of KL-divergence for the Stag Hunt game; in this case the distribution $P(t,\alpha)$ becomes concentrated over time either at $\alpha = 1$ or $\alpha = 0$, depending on the initial distribution.

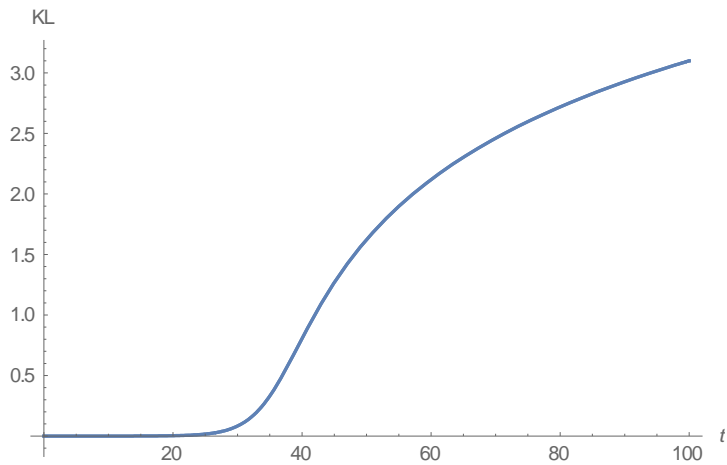

**Figure 12**. Dynamics of KL-divergence for Stag Hunt game: $A = B = 1$, $s = \pm 0.01$. Due to symmetry, dynamics of information gain is identical in both cases.

Figure 13 shows dynamics of KL-divergence for Hawk-Dove game; in this case the limit distribution $P(t, \alpha)$ as $t \to \infty$ is non-singular, and its support coincides with the support of initial distribution. The limit value of $I[P_t : P_0]$ as $t \to \infty$ is finite, in contrast with all other games.

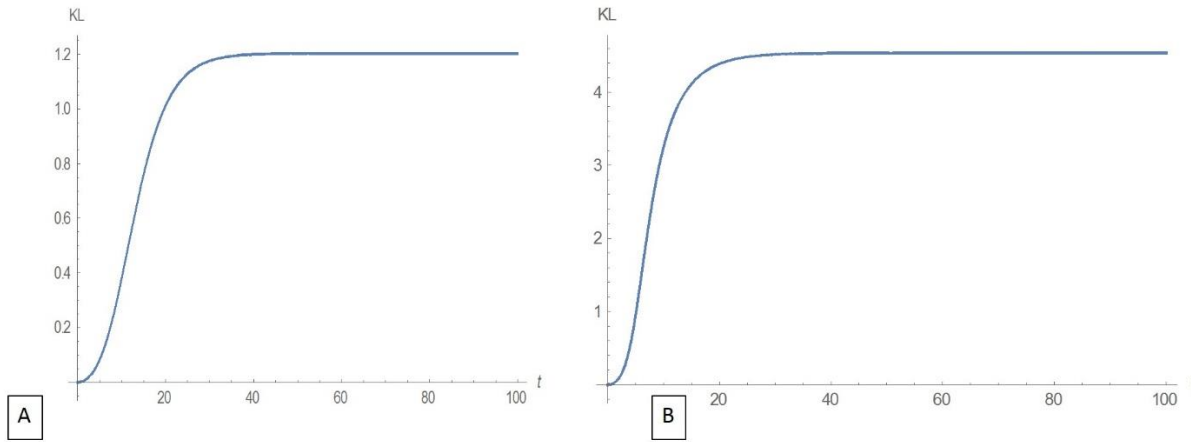

**Figure 13**. Dynamics of KL-divergence for Hawk-Dove game. Left panel: $A = -1, B = 1, s = 10$. Right panel: $A = -1, B = 1, s = -10$.

Overall, within the framework of the considered model of strategy selection, the dynamical version of the Principle of minimum of information gain can be derived from the model's dynamics instead of being postulated *a priori*. Minimization of information gain at any time point is an intrinsic property of the process of natural selection of strategies for all games due to the game dynamics. Initial distribution of strategies and the current mean value, which is equal to the current frequency of one of pure strategies, completely define the current distribution of strategies due to MinxEnt; no other details about the selection process or the game itself are necessary.

The Principle of minimum of information gain is not an external hypothesis but a mathematical theorem within the frameworks of the model of strategy selection. It can therefore be considered an underlying variational principle, which governs the selection process for all games at every time moment.

## Discussion

In this paper, we investigated the process of natural selection of mixed strategies in classical two-player two-strategy games. The current state of the strategy selection process is described by the probability distribution of the parameter $\alpha$, which represents the (heritable) probability of an individual keeping one of two strategies. This distribution can be considered as a distribution of mixed strategies in the game, and its dynamics is the main problem of interest.

The problem was completely solved using the hidden keystone variable (HKV)-method; the distribution of the parameter $\alpha$ at any time moment is defined by formulas (3.2) - (3.4). The dynamics of frequencies of pure strategies is described by the equation (1.7), which generalizes a well-known replicator equation by Taylor& Jonker (1978).

These general results were applied to several known games. We were able to show that the dynamics of strategy distributions in Prisoner's dilemma (PD) and Harmony (H) games essentially depend on the initial distribution of mixed strategies. In both cases the limit distribution is singular, i.e. only a pure strategy can be selected over time from all possible mixed strategies, in accordance to known results.

Similarly, in the Stag Hunt (SH)-game, only one pure strategy can be selected over time but the initial population composition, and more specifically, its mean value, has a critical impact on what strategy will finally be selected. For this reason the result of natural selection of strategies may look like a random process, since it critically depends on (perhaps, random) variations of initial composition of the population.

In all of these cases, only a single pure strategy can be selected; we can interpret these outcomes as "Darwinian selection" of the "fittest" strategy.

In contrast, in the Hawk-Dove (HD) game, not only the overall dynamics but also the shape of final distribution of mixed strategies depends on the initial distribution. The final distribution is not singular, and any mixed strategy that was initially present in the population will be present in the final distribution. We can interpret this outcome as "non-Darwinian survival of everybody." Another principal difference of HD-game from all other games is that KL-divergence between current and initial distribution of strategies (information gain) tends to a finite value, while it tends to infinity for all other games. We would like to emphasize that HD-game has a clear biological interpretation and may be used to explain the persistence of species, whose members have potentially lethal attributes.

Interestingly, the process of natural selection of strategies for all games obeys the dynamical Principle of minimum of information gain. It means that given the initial distribution and the value of frequency of one of pure strategies, at any time moment the current distribution of mixed strategies provides minimum of information gain over *all probability distributions*. Formally, we can postulate this principle (as Kullback and Jaynes did in statistics and statistical physics accordingly) and then construct the solution to the model by solving a corresponding variational problem. What is important is that now we know for certain that this way we obtain the distribution that *exactly coincides* with the solution to the model. Hence, the Principal of minimum of information gain is the underlying *optimization principle*, whose "invisible hand" governs the process of strategy selection in all games.

## Acknowledgements

This research was supported by the Intramural Research Program of the NCBI, NIH.